\documentclass[9pt, conference]{IEEEtran}
\IEEEoverridecommandlockouts

\usepackage{cite}
\usepackage{amsmath,amssymb,amsfonts}
\usepackage{algorithmic}
\usepackage{graphicx}
\usepackage{textcomp}
\usepackage[hidelinks]{hyperref}
\usepackage{xcolor}
\def\BibTeX{{\rm B\kern-.05em{\sc i\kern-.025em b}\kern-.08em
    T\kern-.1667em\lower.7ex\hbox{E}\kern-.125emX}}

\def\name#1{\gdef\@name{{\em #1}\\}}
\def\address#1{\gdef\@address{#1}}

\begin{document}

\title{BEST-STD: Bidirectional Mamba-Enhanced Speech Tokenization for Spoken Term Detection\\
\thanks{This work was supported by a research grant from MeitY, GoI.}
}

\author{
    \IEEEauthorblockN{\textit{Anup Singh}\IEEEauthorrefmark{1}\IEEEauthorrefmark{2}, \textit{Kris Demuynck}\IEEEauthorrefmark{1}, \textit{Vipul Arora}\IEEEauthorrefmark{2}}
    \IEEEauthorblockA{\IEEEauthorrefmark{2}Department of Electrical Engineering, Indian Institute of Technology Kanpur, India}
    \IEEEauthorblockA{\IEEEauthorrefmark{1}IDLab, Department of Electronics and Information Systems, imec - Ghent University, Belgium
    \\\{anup.singh, kris.demuynck\}@ugent.be, vipular@iitk.ac.in}
}

\maketitle

\begin{abstract}
Query-by-example spoken term detection (QbE-STD) is often hindered by reliance on frame-level features and the computationally intensive DTW-based template matching, limiting its practicality. To address these challenges, we propose a novel approach that encodes speech into discrete, speaker-agnostic semantic tokens. This facilitates fast retrieval using text-based search algorithms and effectively handles out-of-vocabulary terms. Our approach focuses on generating consistent token sequences across varying utterances of the same term. We also  propose a bidirectional state space modeling within the Mamba encoder, trained in a self-supervised learning framework, to learn contextual frame-level features that are further encoded into discrete tokens. Our analysis shows that our speech tokens exhibit greater speaker invariance than those from existing tokenizers, making them more suitable for QbE-STD tasks. Empirical evaluation on LibriSpeech and TIMIT databases indicates that our method outperforms existing baselines while being more efficient.
\end{abstract}


\section{Introduction}
 Query-by-example spoken term detection (QbE-STD) involves retreiving utterances within large audio archives that contains a given spoken query. It is useful in various applications \cite{b1,b2}, including voice search \cite{b3} in multimedia content, particularly for broadcast programs, lectures, meeetings, etc.

The traditional ASR-based approaches in STD represent spoken content using subwords units such as phone or grapheme lattices \cite{b4,b5,b6} to address out-of-vocabulary (OOV) problems. However, these methods rely on highly accurate ASR systems, which are challenging to develop, especially for short-duration queries spoken without context and under varying acoustic conditions \cite{b7,b8}. To mitigate these challenges, \cite{b9, b10} rely on direct audio signal comparison via template matching techniques like segmental-dynamic time warping (SDTW)\cite{b11}. However, performing an exhaustive search in a large database with computationally expensive SDTW renders these STD systems impractical. An alternative approach to STD involves learning discriminative acoustic word embeddings \cite{b12, b13, b14, b15, b16_0}. A major drawback of this approach is the difficulty in segmenting words within continuous speech, where word boundaries are often indistinguishable. To address this challenge, recent audio fingerprinting methods \cite{b16_1, b16_2, b16_3} offer an alternative by eliminating the need for explicit word boundary detection; however, their effectiveness is limited by a lack of robustness to speaker variation. Lastly, current evaluation methods for STD systems typically test query-segment pairs to detect query presence \cite{b9, b10, b11, b12, b13, b14, b15, b16_0}, which does not reflect real-world performance, where the goal is to retrieve an ordered list of spoken documents containing the query.

To address these limitations, we propose employing speech tokenization using discrete speech representations. This approach  offers several advantages: it enables the use of efficient text-based search algorithms, eliminates the need for explicit word segmentation, allows for compact storage of speech data as token sequences, and effectively handles OOV words by learning subword units as discrete tokens.

In this paper, we propose a speech tokenizer, dubbed \textbf{BEST-STD}, that generates speaker-agnostic speech tokens and captures subword information. Our approach first transforms speech into a sequence of contextual embeddings with our novel bidirectional Mamba encoder. These embeddings are then discretized into a sequence of tokens. Our model is trained within a self-supervised learning framework to generate consistent token sequences across different utterances of the same spoken term. For instance, the model tokenizes two different utterances of the term “\textit{hello}” as $\{1, 1, 3, 3, 8\}$ and $\{1, 3, 3, 3, 8, 8\}$. We focus on generating discriminative frame-level embeddings instead of word-level embeddings. During training, we leverage DTW to align utterances of the same word, constructing anchor-positive pairs at the frame level. To efficiently retrieve utterances containing a spoken term, we use an inverted index to index the tokenized speech archive. We evaluate our method on the LibriSpeech \cite{b16} and TIMIT\cite{b17} databases, demonstrating its superior performance compared to existing STD baselines. Furthermore, our analysis shows that our method generates more discriminative token sequences than those from existing speech tokenizers\cite{b18, b19, b20, b21}.

\section{Method}

\subsection{Representation Learning}\label{AA}
\textbf{Mathematical framework of the Mamba model.} Structured State-Space Sequence models (S4) \cite{b22} are linear time-invariant models derived from continuous system principles. These models map a multivariate input sequence $x(t) \in \mathbb{R}^D$ to an output sequence $y(t) \in \mathbb{R}^D$ through a hidden state $h(t) \in \mathbb{R}^{N}$. This mapping is governed by the evolution parameters $\mathbf{A} \in \mathbb{R}^{N\times N}$ and projection parameters $\mathbf{B} \in \mathbb{R}^{N\times D}$ and $\mathbf{C} \in \mathbb{R}^{D\times N}$, following the differential equations:
\begin{equation}
    \begin{aligned}
        h'(t) &= \mathbf{A}h(t) + \mathbf{B}x(t), \quad y(t) &= \mathbf{C}h(t)
    \end{aligned}
    \label{cts-differntial}
\end{equation}

Here, $\mathbf{A}$ is typically initialized with a HiPPO matrix \cite{hippo}  or a Diagonal matrix \cite{diagonal} to capture long-term dependencies effectively. These continuous equations can be discretized by introducing a timescale parameter $\mathbf{\Delta}$, transforming the continuous matrices $\mathbf{A}$ and $\mathbf{B}$ into their discrete counterparts $\mathbf{\Tilde{A}}$ and $\mathbf{\Tilde{B}}$. This transformation typically employs the Zero-Order Hold (ZOH) method as: 
\begin{equation}
    \begin{aligned}
    \mathbf{\Tilde{A}} &= \exp (\Delta \mathbf{A}), \quad \mathbf{\Tilde{B}} &= (\Delta \mathbf{A})^{-1} \left( \exp (\Delta \mathbf{A}) - \mathbf{I} \right) \cdot \Delta \mathbf{B}.
    \end{aligned}
\end{equation}

The discretized version of \eqref{cts-differntial} can then be expressed as:
\begin{equation}
    \begin{aligned}
        h_t &= \mathbf{\Tilde{A}} h_{t-1} + \mathbf{\Tilde{B}} x_t, \quad y_t &= \mathbf{C} h_t.
    \end{aligned}
\end{equation}



Mamba \cite{b23} enhances this framework by converting its time-invariant parameters into time-variant parameters, denoted as $\mathbf{A}_t$, $\mathbf{B}_t$, $\mathbf{C}_t$, and $\mathbf{\Delta}_t$. These parameters are dynamically updated at each timestep $t$ based on the input $x_t$, allowing the model to adaptively evolve into a \textit{selective} structured state-space model.

\begin{figure}[t]
    \centering
    \includegraphics[width=\linewidth]{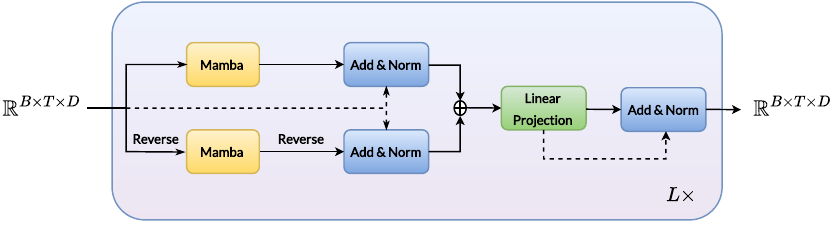}
    \caption{The architecture of proposed bidirectional Mamba encoder.}
    \label{bimamba}
\end{figure}

\textbf{Bidirectional Mamba Encoder.} The original Mamba model operates unidirectionally, limiting its ability to capture contextual information effectively. To address this, we propose a straightforward bidirectional modeling approach that employs two structurally identical Mamba blocks to capture long-range dependencies independently in both forward and backward directions. The forward Mamba block processes the input sequence, while the backward Mamba block simultaneously processes a time-reversed version of the input. The output from the backward block is then reversed back and combined with the output of the forward block. This combined output is subsequently passed through an output projection layer, as illustrated in Figure \ref{bimamba}. Moreover, the output from the last bidrectional Mamba layer is projected into a $d$-dimensional space and $L_2$-normalized to produce the final encoder output.

\begin{figure*}
    \centering
    \includegraphics[width=\textwidth]{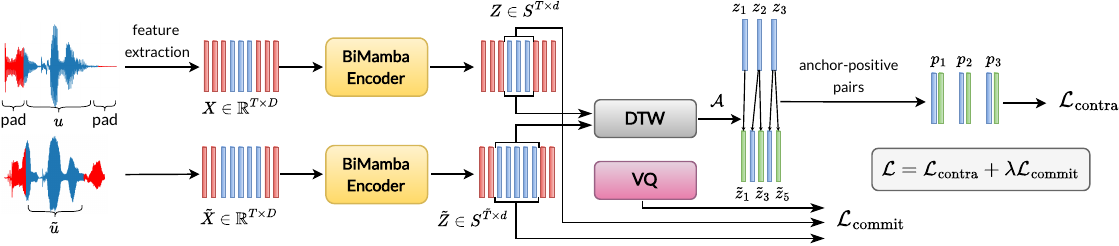}
    \caption{Illustration of our self supervised learning framework for learning speech tokens.}
    \label{system_overview}
\end{figure*}

\subsection{Self-Supervised Learning Framework} 
Our objective is to train a model that generates consistent token sequences $Z_q$ and $\Tilde{Z}_q$ for pairs of utterances ($u, \Tilde{u}$) corresponding to a spoken term $w$. An overview of our learning framework is illustrated in Figure \ref{system_overview}.

\textbf{Discriminative frame-level embeddings.} We first consider a pair of utterances with durations $t$ and $\Tilde{t}\;(\textrm{WLOG}, \Tilde t\geq t)$ for a term $w$. To better simulate real-world scenarios where spoken queries may not be isolated, we add contextual padding to these utterances to create fixed-length audio segments. These segments are processed to extract frame-level acoustic features, such as log Mel spectrum or MFCCs, resulting in sequences $X$ and $\Tilde{X}$. These sequences are then input to the encoder to generate embedding sequences $Z$ and $\Tilde{Z}$. When computing loss during training, we exclude embeddings corresponding to the padding, yielding sequences $Z=\{z_t\}_{t=1}^T$ and $\Tilde{Z}=\{\Tilde{z}_t\}_{t=1}^{\Tilde{T}}$.  Next, we obtain the DTW alignment $\mathcal{A}$ between $Z$ and $\Tilde{Z}$ as: 
\begin{equation}
    \mathcal{A} = \{(t, S_t) \mid t \in [1, T], S_t \subseteq [1, \Tilde{T}]\},
    \label{alignment}
\end{equation}

where $t$ denotes the frame index in $Z$ and $S_t$ the set of indices in $\Tilde{Z}$ aligned with $t$. We then leverage $\mathcal{A}$ to generate anchor-positive pairs $p_t$ at the frame-level:
\begin{equation}
    p_t = (z_t, \Tilde{z}_{t^*}), \;\textrm{where}\; t^*=\arg \max_{j \in S_t} \cos({z_{i}\cdot \Tilde{z}_{j}}), 
\end{equation}
where 
$t^*$ is the frame index in $\Tilde{Z}$ that maximizes the cosine similarity with $z_t$.

Hence, for each training pair ($Z, \Tilde{Z}$), indexed by $i$, we define the contrastive loss function as:
\begin{equation}
    \mathcal{L}_{\textrm{contrast}}^{(i)} = \frac{1}{T} \sum\limits_{t=1}^T -\log \Bigg(\frac{e^{(z_t \cdot \Tilde{z}_{t^*}/\tau)}}{e^{(z_t \cdot \Tilde{z}_{t^*}/\tau)}+ \sum\limits_{n=1}^{N} e^{(z_t \cdot z_{n}/\tau)}}\Bigg),
\label{contra_loss}
\end{equation}

where $z_n$ are $N$ negative embeddings randomly chosen from other training pairs in a batch, i.e., embedding from term $w' \neq w$.

\textbf{Tokenization.} Finally, we quantize each embedding in $Z$ and $\Tilde{Z}$ using a  vector quantizer (VQ) $h(.)$ to obtain their corresponding discrete token sequences $Z_q$ and $\Tilde{Z}_q$ as:
\begin{equation}
    \begin{aligned}
        z_q = h(z) = \arg \min_{c_i \in C} \left \Vert z-c_i \right \Vert_2^2,
    \end{aligned}
\end{equation}

where $C=\{c_1, ..., c_K\}$ is a set of $d$-dimensional cluster centroids. We implicitly optimize the codebook $C$ using exponential moving average updates \cite{b24} and thus avoid the instability and potential codebook collapse associated with codebook loss. We also apply $L_2$ normalization to centroids to enhance training stability and improve codebook utilization\cite{b25}.

To further encourage the encoder to generate embeddings aligned with the cluster centroids, we introduce a commitment loss function for the $i^{th}$ pair as:
\begin{equation}
    \mathcal{L}_{\textrm{commit}}^{(i)} = \frac{1}{T}\sum\limits_{t=1}^T\left \Vert z_t-z_{q_{t}} \right \Vert_2^2 + \frac{1}{\Tilde{T}}  \sum\limits_{t=1}^{\Tilde{T}}\left \Vert \Tilde{z}_t-\Tilde{z}_{q_{t}} \right \Vert_2^2
\end{equation}

\textbf{Loss function.} Finally, the total loss $\mathcal{L}$ over the training batch of size $B$ is defined as:
\begin{equation}
    \mathcal{L} = \frac{1}{B} \sum\limits_{i=1}^B \mathcal{L}_{\textrm{contrast}}^{(i)} + \lambda \mathcal{L}_{\textrm{commit}}^{(i)},
    \label{total_loss}
\end{equation}
where $\lambda\geq0$ controls tradeoff between the two loss components.

\subsection{Indexing}
To facilitate efficient retrieval of spoken documents, we construct an inverted index $I$ by processing each audio track in the database as follows. Each track $a_i$ is divided into overlapping segments, indexed by $j$, of length $l$ with a hop interval $h$. Each segment $s_{ij}$ is then tokenized into a sequence $Z_{ij}=\{z_1, .., z_T\}$, where each token $z_t \in \{1, ..., K\}$. 

Since we compare token sequences using Jaccard similarity \cite{jaccard}, which does not inherently account for temporal information, we generate bigrams from the extracted tokens. This results in a bigram sequence $B_{ij}=\{b_1, b_2, ..., b_{T-1}\}$, which reflect the local token order and thereby encodes some temporal structure. As such, we ensure the temporal information is incorporated when computing similarity between segments.

The inverted index $I$ is then constructed with each bigram $b_k$ as a key that maps to a list of pairs $(i,j)$,  such that $b_k \in B_{ij}$:
\begin{equation}
    I(b_k) = \{(i,j)| b_k \in B_{ij}\}, b_k \in \{1, ..., K\} \times \{1, ..., K\}
\end{equation}

\subsection{Retrieval}

For a given spoken term query $q$, we tokenize it to generate a bigram token sequence $Q = \{b_1,b_2, ..., b_{\Tilde{T}}\}$. We first perform a coarse search to identify the best matching frame candidates by collecting frames that contain at least one query bigram:
\begin{equation}
    C = \bigcup_{b_k \in Q} I(b_k)
\end{equation}

Next, we perform a fine-grained search to precisely locate the spoken query within each candidate frame $c_k \in C$. For each frame $c_k$, we compute the maximum Jaccard similarity between $Q$ and every possible subsequence of $c_k$ of length $\Tilde{T}$, starting at index $t$:
\begin{equation}
    s(c_k) = \max_{t \in \{1,2,..., T-\Tilde{T}+1\}} J(Q, c_{k_{[t:t+\Tilde{T}-1]}})
\end{equation}

Finally, we rank all candidate frames $c_k \in C$ based on their Jaccard similarity $s(c_k)$ and select the top-$k$ candidates.

\section{Experimental Setup}

\subsection{Databases}
We employed the LibriSpeech train-clean-360 subset to train our model. We utilized the test-clean subset as a validation set to optimize model parameters. Additionally, the train-clean-100 subset, containing approximately 28k utterances and totaling around 100 hours of spoken content, served as our speech archive. For testing, we extracted spoken terms from the train-clean-100 set, ensuring no overlap between the training and testing sets. We created two distinct query sets, each consisting of 300 unique terms:

\begin{itemize}
    \item Query Set 1: This set consists of in-vocabulary (IV) terms—queries whose text forms appeared during training, but are uttered by speakers not seen during training.
    \item Query Set 2: This set comprises out-of-vocabulary (OOV) terms—queries whose text forms, as well as the speakers, were not seen during training.
\end{itemize}

\subsection{Evaluation Metrics}
We used the Mean Reciprocal Rank (MRR)\cite{map}, the Mean Average Precision (MAP) \cite{map}, and the Maximum Term Weighted Value (MTWV) \cite{mtwv} metrics to evaluate all methods. 

\subsection{Baselines}
We compared our proposed method with existing STD approaches that generate frame-level features, such as MFCCs \cite{b15}, bottleneck features (BNF) \cite{b26}, and posterior probabilities \cite{b27}, followed by DTW-based template matching. Also, we evaluated our method against established speech tokenizers, including HuBERT\cite{b19}, WavLM\cite{b20}, SpeechTokenizer\cite{b21}, and EnCodec\cite{b22} to assess their effectiveness in the STD task. The KMeans models for HuBERT and WavLM were extracted from the Speechbrain toolkit \cite{bsb}, and each model was trained on speech representations extracted from the seventh layer of its respective model.

\subsection{Implementation details}
We extracted spoken utterances from the audio tracks and applied contextual padding to create 1-second ($l$) audio segments.. This fixed input length was determined based on a statistical analysis of the LibriSpeech train-clean-360 subset, which revealed that approximately 93\% of unique terms had utterance durations of less than 1 second. By selecting this 1-second length, we ensured that the majority of unique terms were effectively utilized during training.

Each audio segment was then converted into a 96-dimensional ($D$) Mel-spectrogram, which served as the input to the encoder. The encoder architecture consisted of four bidirectional Mamba layers, followed by a feedforward network that projected the final layer output into a 512-dimensional ($d$) embedding space. In total, the model contained 4.7M trainable parameters. We experimented with varying codebook sizes ($K$), ranging from 128 to 1024, to assess their impact on the system performance. The hyperparameter $\tau$ in (\ref{contra_loss}) was fixed at 0.2 and $\lambda$ in (\ref{total_loss}) was set to 0.1. The model was trained for 600 epochs using the Adam optimizer with a learning rate set to 5e-4.

\section{Results}

\begin{table}[]
\centering
\caption{The average Jaccard similarity between discrete representations of pairs of spoken term utterances.}
\begin{tabular}{|c|c|cc|}
\hline
\textbf{Tokenizer} & \textbf{Tokens} & \textbf{Unigram} & \textbf{Bigram} \\ \hline
HuBERT-Base        & 512             & 0.31             & 0.11            \\
HuBERT-Base        & 1000            & 0.26             & 0.14            \\
WavLM-Base         & 512             & 0.40             & 0.21            \\
WavLM-Base         & 1000            & 0.33             & 0.19            \\
Encodec            & 1024            & 0.16             & 0.08            \\
SpeechTokenizer    & 1024            & 0.51             & 0.31            \\ \hline
Ours:              &                 &                  &                 \\
Transformer        & 512             & 0.74             & 0.64            \\
Transformer        & 1024            & 0.71             & 0.60            \\
BEST-STD       & 256             & \textbf{0.84}             & \textbf{0.77}            \\
BEST-STD       & 512             & 0.80             & 0.72            \\
BEST-STD       & 1024            & 0.78             & 0.69            \\ \hline
\end{tabular}
\label{jaccard_sim}
\end{table}

\begin{table*}[ht]
\centering
\caption{Spoken content retrieval results (higher the better) on LibriSpeech train-clean-100 subset and TIMIT datasets.}
\begin{tabular}{|c|c|cccccc|cccccc|}
\hline
                       &                 & \multicolumn{6}{c|}{\textbf{LibriSpeech}}                                                                          & \multicolumn{6}{c|}{\textbf{TIMIT}}                                                                                \\ \hline
                       &                 & \multicolumn{3}{c|}{\textbf{In-Vocabulary}}                      & \multicolumn{3}{c|}{\textbf{Out-of-Vocabulary}} & \multicolumn{3}{c|}{\textbf{In-Vocabulary}}                      & \multicolumn{3}{c|}{\textbf{Out-of-Vocabulary}} \\ \hline
\textbf{Methods}       & \textbf{Tokens} & \textbf{MAP} & \textbf{MRR} & \multicolumn{1}{c|}{\textbf{MTWV}} & \textbf{MAP}   & \textbf{MRR}  & \textbf{MTWV}  & \textbf{MAP} & \textbf{MRR} & \multicolumn{1}{c|}{\textbf{MTWV}} & \textbf{MAP}   & \textbf{MRR}  & \textbf{MTWV}  \\ \hline
MFCC             & -               & 0.32         & 0.37         & \multicolumn{1}{c|}{0.48}          & 0.41           & 0.46          & 0.45           & 0.31         & 0.39         & \multicolumn{1}{c|}{0.50}          & 0.45           & 0.46          & 0.44           \\
Phone Posteriors & -               & 0.44         & 0.46         & \multicolumn{1}{c|}{0.53}          & 0.49           & 0.53          & 0.51           & 0.43         & 0.46         & \multicolumn{1}{c|}{0.55}          & 0.43           & 0.45          & 0.49           \\
BNF              & -               & 0.16         & 0.26         & \multicolumn{1}{c|}{0.18}          & 0.17           & 0.20          & 0.12           & 0.20         & 0.28         & \multicolumn{1}{c|}{0.20}          & 0.22           & 0.25          & 0.24           \\
HuBERT-Base            & 512             & 0.29         & 0.32         & \multicolumn{1}{c|}{0.52}          & 0.29           & 0.30          & 0.66           & 0.26         & 0.26         & \multicolumn{1}{c|}{0.42}          & 0.33           & 0.34          & 0.40           \\
HuBERT-Base            & 1000            & 0.23         & 0.26         & \multicolumn{1}{c|}{0.42}          & 0.28           & 0.27          & 0.40           & 0.24         & 0.22         & \multicolumn{1}{c|}{0.30}          & 0.28           & 0.28          & 0.21           \\
WavLM-Base             & 512             & 0.44         & 0.49         & \multicolumn{1}{c|}{0.57}          & 0.44           & 0.45          & 0.60           & 0.38         & 0.38         & \multicolumn{1}{c|}{0.48}          & 0.40           & 0.41          & 0.44           \\
WavLM-Base             & 1000            & 0.38         & 0.39         & \multicolumn{1}{c|}{0.49}          & 0.40           & 0.37          & 0.47           & 0.33         & 0.34         & \multicolumn{1}{c|}{0.38}          & 0.37           & 0.37          & 0.39           \\
Encodec                & 1024            & 0.20         & 0.21         & \multicolumn{1}{c|}{0.31}          & 0.21           & 0.21          & 0.35           & 0.10         & 0.11         & \multicolumn{1}{c|}{0.17}          & 0.04           & 0.03          & 0.21           \\
SpeechTokenizer        & 1024            & 0.57         & 0.62         & \multicolumn{1}{c|}{0.56}          & 0.53           & 0.53          & 0.65           & 0.46         & 0.48         & \multicolumn{1}{c|}{0.46}          & 0.45           & 0.46          & 0.45           \\ \hline
\textbf{Ours:}         &                 &              &              & \multicolumn{1}{c|}{}              &                &               &                &              &              & \multicolumn{1}{c|}{}              &                &               &                \\
Transformer           & 512             & 0.80         & 0.84         & \multicolumn{1}{c|}{0.63}          & 0.76           & 0.77          & 0.56           & 0.69         & 0.74         & \multicolumn{1}{c|}{0.76}          & 0.66           & 0.73          & 0.69           \\
Transformer           & 1024            & 0.77         & 0.82         & \multicolumn{1}{c|}{0.68}          & 0.73           & 0.74          & 0.61           & 0.66         & 0.73         & \multicolumn{1}{c|}{0.70}          & 0.65           & 0.69          & 0.64           \\
BEST-STD               & 256             & \textbf{0.86}         & 0.90         & \multicolumn{1}{c|}{0.62}          & \textbf{0.83}           & \textbf{0.84}          & 0.55           & \textbf{0.75}         & \textbf{0.78}         & \multicolumn{1}{c|}{0.69}          & \textbf{0.70}           & \textbf{0.75}          & 0.63           \\
BEST-STD               & 512             & 0.86         & \textbf{0.91}         & \multicolumn{1}{c|}{0.66}          & 0.82           & 0.83          & 0.60           & 0.72         & 0.78         & \multicolumn{1}{c|}{0.74}          & 0.69           & 0.75          & 0.65           \\
BEST-STD               & 1024            & 0.78         & 0.84         & \multicolumn{1}{c|}{\textbf{0.73}}          & 0.77           & 0.78          & \textbf{0.65}           & 0.68         & 0.75         & \multicolumn{1}{c|}{\textbf{0.75}}          & 0.66           & 0.71          & \textbf{0.70}           \\ \hline
\end{tabular}
\label{main_results}
\end{table*}

\subsection{Analysis of discrete representations}

First, we aim to validate that our method consistently generates discrete representations for different utterances of the same spoken term. To this end, we randomly selected 1,000 unique terms from the LibriSpeech corpus, extracting pairs of corresponding utterances and encoding them into discrete token sequences. We then evaluated the consistency of these sequences using the Jaccard similarity metric, considering both unigrams and bigrams as tokens. 

As presented in Table \ref{jaccard_sim}, our tokenizer outperforms state-of-the-art speech tokenizers, achieving significantly higher Jaccard similarity scores across varying numbers of cluster centroids. These results highlight the speaker-agnostic nature of our discrete tokens, demonstrating their effectiveness for STD tasks. Furthermore, despite being trained within the same framework, the bidirectional Mamba outperforms the Transformer. We attribute this difference to the positional encoding employed in Transformers, which may inadequately capture unwanted temporal details. In contrast, the bidirectional Mamba model relies on a simpler temporal model, which is beneficial in the STD context. 

Using a smaller number of centroids, such as 256, our method achieves the highest Jaccard similarity score. However, this configuration diminishes discriminative power, as it tokenizes phonetically close words into identical discrete representations. To address this issue, it is crucial to increase the number of centroids that capture finer subword units, thereby enhancing the discriminative power of the model.

\subsection{Spoken content retrieval}

Table \ref{main_results} depicts how the proposed method performs against the existing speech tokenizers and conventional STD baselines for spoken content retrieval tasks. The results show that our method consistently outperforms all baselines across both test sets. We observe that our method particularly excels in detecting IV terms and effectively handles OOV terms. This is because our tokens capture subword information such as syllables and exhibit more robustness against speaker variations than those from the baseline tokenizers. Consequently, our method enables more accurate and reliable retrieval.  Among the baseline tokenizers, SpeechTokenizer showed the highest performance due to its generation of semantic tokens. In contrast, Encodec performed the worst since it generates acoustic tokens that still encode speaker information, leading to diminished retrieval accuracy.  It is also important to highlight that, while the baseline tokenizers were pre-trained on the entire LibriSpeech corpus --- meaning the query terms in both query sets were likely encountered during training --- our model achieved superior results despite being trained on a smaller subset with significantly fewer parameters. \\
The results further reveal that the proposed method outperforms traditional DTW-based baselines, underscoring the effectiveness of our approach. Specifically, the use of robust semantic speech tokens proves sufficient for comparing speech segments, eliminating the need for continuous speech representations. Additionally, by leveraging an inverted index, our method delivers significantly faster retrieval times compared to DTW-based methods.  Overall, these findings highlight the effectiveness and efficiency of speech tokenization in handling STD tasks, reinforcing the practical applicability of the proposed method.\\
The comparison between the Transformer model and the bidirectional Mamba model reveals that the latter demonstrates superior performance. The bidirectional Mamba achieves this by generating more discriminative discrete representations of spoken terms primarily due to its ability to better model fine-grained local temporal information, which is more important for the STD task than the global temporal information at which transformers excel. \\
Our ablation study on the impact of codebook sizes demonstrates that increasing the number of centroids leads to an improvement in MTWV scores. This indicates that larger codebooks capture finer discrete units, thereby improving the discriminability of spoken terms. However, we observed a slight decline in MAP and MRR scores with larger codebook sizes. These findings highlight the trade-off between different performance metrics, suggesting that the selection of codebook size should be based on the specific use case.

We also found that our method occasionally generated false positives during retrieval, particularly with homophones, such as "plane" vs "plain," which were tokenized into identical token sequences. This performance drop highlights the need for further investigation into challenges posed by phonetically similar words.

\section{Conclusion and Future Work}
In this paper, we introduced a novel approach to STD by encoding speech into discrete tokens. We also proposed a bidirectional state space modeling using the Mamba model to effectively learn contextual speech representations. Extensive experiments demonstrate that our approach outperforms existing STD baselines in accuracy and efficiency. Further analysis reveals that our method generates tokens with greater invariance to speaker variability, making them more suitable for STD tasks than current speech tokenizers. Moreover, the bidirectional Mamba model surpasses the Transformer model due to better modeling of temporal information. Our speech tokenization framework also shows promise for applications in developing speech LLMs \cite{b28,b29,b30,b31}. Future work will focus on generating language-agnostic speech tokens and improving retrieval efficiency. Our code is available at: \href{https://github.com/anupsingh15/BEST-STD}{\texttt{https://github.com/anupsingh15/BEST-STD}}


\begin{thebibliography}{00}




\bibitem{b1} Alberti, Christopher, Michiel Bacchiani, Ari Bezman, Ciprian Chelba, Anastassia Drofa, Hank Liao, Pedro Moreno et al. "An audio indexing system for election video material." In 2009 IEEE International Conference on Acoustics, Speech and Signal Processing, pp. 4873-4876. IEEE, 2009.

\bibitem{b2} Ogata, J. and Goto, M., 2009. PodCastle: Collaborative training of acoustic models on the basis of wisdom of crowds for podcast transcription. In Tenth Annual Conference of the International Speech Communication Association.

\bibitem{b3} Wang, Y.Y., Yu, D., Ju, Y.C. and Acero, A., 2008. An introduction to voice search. IEEE Signal Processing Magazine, 25(3), pp.28-38.


\bibitem{b4} Mamou, J., Ramabhadran, B. and Siohan, O., 2007, July. Vocabulary independent spoken term detection. In Proceedings of the 30th annual international ACM SIGIR conference on Research and development in information retrieval (pp. 615-622).

\bibitem{b5} Miller, D.R., Kleber, M., Kao, C.L., Kimball, O., Colthurst, T., Lowe, S.A., Schwartz, R.M. and Gish, H., 2007, August. Rapid and accurate spoken term detection. In Interspeech (Vol. 7, pp. 314-317).

\bibitem{b6}Wang, D., Frankel, J., Tejedor, J. and King, S., 2008, March. A comparison of phone and grapheme-based spoken term detection. In 2008 IEEE International Conference on Acoustics, Speech and Signal Processing (pp. 4969-4972). IEEE.

\bibitem{b7} Saraclar, M. and Sproat, R., 2004. Lattice-based search for spoken utterance retrieval. In Proceedings of the Human Language Technology Conference of the North American Chapter of the Association for Computational Linguistics: HLT-NAACL 2004 (pp. 129-136).

\bibitem{b8} Can, D. and Saraclar, M., 2011. Lattice indexing for spoken term detection. IEEE Transactions on Audio, Speech, and Language Processing, 19(8), pp.2338-2347.

\bibitem{b9} Ram, D., Miculicich, L. and Bourlard, H., 2020. Neural network based end-to-end query by example spoken term detection. IEEE/ACM Transactions on Audio, Speech, and Language Processing, 28, pp.1416-1427.

\bibitem{b10} Ram, D., Miculicich, L. and Bourlard, H., 2018, September. CNN Based Query by Example Spoken Term Detection. In Interspeech (pp. 92-96).

\bibitem{b11} Tsai, T.J., 2021, June. Segmental DTW: A parallelizable alternative to dynamic time warping. In ICASSP 2021-2021 IEEE International Conference on Acoustics, Speech and Signal Processing (ICASSP) (pp. 106-110). IEEE.

\bibitem{b12} Chung, Y.A., Wu, C.C., Shen, C.H., Lee, H.Y. and Lee, L.S., 2016. Unsupervised learning of audio segment representations using sequence-to-sequence recurrent neural networks. In Proc. Interspeech (pp. 765-769).

\bibitem{b13} He, W., Wang, W. and Livescu, K., 2016. Multi-view recurrent neural acoustic word embeddings. arXiv preprint arXiv:1611.04496.

\bibitem{b14} Chen, Y.C., Huang, S.F., Shen, C.H., Lee, H.Y. and Lee, L.S., 2018, December. Phonetic-and-semantic embedding of spoken words with applications in spoken content retrieval. In 2018 IEEE Spoken Language Technology Workshop (SLT) (pp. 941-948). IEEE.

\bibitem{b15} Kamper, H., Wang, W. and Livescu, K., 2016, March. Deep convolutional acoustic word embeddings using word-pair side information. In 2016 IEEE International Conference on Acoustics, Speech and Signal Processing (ICASSP) (pp. 4950-4954). IEEE.

\bibitem{b16_0} Hu, Y., Settle, S., and Livescu, K., 2021, January. Acoustic span embeddings for multilingual query-by-example search. In 2021 IEEE Spoken Language Technology Workshop (SLT) (pp. 935-942). IEEE.

\bibitem{b16_1} Singh, A., Demuynck, K. and Arora, V., 2022. Attention-based audio embeddings for query-by-example. In Proceedings of the 23rd International Society for Music Information Retrieval Conference, {ISMIR} 2022 (pp. 52-58)

\bibitem{b16_2} Singh, A., Demuynck, K., and Arora, V., 2024. FlowHash: Accelerating Audio Search with Balanced Hashing via Normalizing Flow. IEEE/ACM Transactions on Audio, Speech, and Language Processing.

\bibitem{b16_3} Singh, A., Demuynck, K., and Arora, V., 2023. Simultaneously learning robust audio embeddings and balanced hash codes for query-by-example. In ICASSP 2023-2023 IEEE International Conference on Acoustics, Speech and Signal Processing (ICASSP) (pp. 1-5). IEEE.



\bibitem{b16} Panayotov, V., Chen, G., Povey, D. and Khudanpur, S., 2015, April. Librispeech: an asr corpus based on public domain audio books. In 2015 IEEE international conference on acoustics, speech and signal processing (ICASSP) (pp. 5206-5210). IEEE.

\bibitem{b17} Garofolo, J.S., Lamel, L.F., Fisher, W.M., Fiscus, J.G. and Pallett, D.S., 1993. DARPA TIMIT acoustic-phonetic continous speech corpus CD-ROM. NIST speech disc 1-1.1. NASA STI/Recon technical report n, 93, p.27403.

\bibitem{b18} Hsu, W.N., Bolte, B., Tsai, Y.H.H., Lakhotia, K., Salakhutdinov, R. and Mohamed, A., 2021. Hubert: Self-supervised speech representation learning by masked prediction of hidden units. IEEE/ACM transactions on audio, speech, and language processing, 29, pp.3451-3460.

\bibitem{b19} Chen, S., Wang, C., Chen, Z., Wu, Y., Liu, S., Chen, Z., Li, J., Kanda, N., Yoshioka, T., Xiao, X. and Wu, J., 2022. Wavlm: Large-scale self-supervised pre-training for full stack speech processing. IEEE Journal of Selected Topics in Signal Processing, 16(6), pp.1505-1518. 

\bibitem{b20} Zhang, X., Zhang, D., Li, S., Zhou, Y. and Qiu, X., 2023. Speechtokenizer: Unified speech tokenizer for speech large language models. arXiv preprint arXiv:2308.16692.

\bibitem{b21} Défossez, A., Copet, J., Synnaeve, G. and Adi, Y., 2022. High fidelity neural audio compression. arXiv preprint arXiv:2210.13438.

\bibitem{b22} Gu, A., Goel, K. and Ré, C., 2021. Efficiently modeling long sequences with structured state spaces. arXiv preprint arXiv:2111.00396.

\bibitem{hippo} Gu, A., Dao, T., Ermon, S., Rudra, A. and Ré, C., 2020. Hippo: Recurrent memory with optimal polynomial projections. Advances in neural information processing systems, 33, pp.1474-1487.

\bibitem{diagonal} Gupta, A., Gu, A. and Berant, J., 2022. Diagonal state spaces are as effective as structured state spaces. Advances in Neural Information Processing Systems, 35, pp.22982-22994.

\bibitem{b23} Dao, T. and Gu, A., 2024. Transformers are SSMs: Generalized models and efficient algorithms through structured state space duality. arXiv preprint arXiv:2405.21060.

\bibitem{b24} Van Den Oord, A. and Vinyals, O., 2017. Neural discrete representation learning. Advances in neural information processing systems, 30.

\bibitem{b25} Yu, J., Li, X., Koh, J.Y., Zhang, H., Pang, R., Qin, J., Ku, A., Xu, Y., Baldridge, J. and Wu, Y., 2021. Vector-quantized image modeling with improved vqgan. arXiv preprint arXiv:2110.04627.

\bibitem{jaccard} Murphy, A.H., 1996. The Finley affair: A signal event in the history of forecast verification. Weather and forecasting, 11(1), pp.3-20.

\bibitem{map} Liu, T.Y., 2009. Learning to rank for information retrieval. Foundations and Trends® in Information Retrieval, 3(3), pp.225-331.

\bibitem{mtwv} Rodriguez-Fuentes, L.J. and Penagarikano, M., 2013. MediaEval 2013 spoken web search task: system performance measures. n. TR-2013-1, Department of Electricity and Electronics, University of the Basque Country.

\bibitem{b26} Silnova, A., Matejka, P., Glembek, O., Plchot, O., Novotný, O., Grezl, F., Schwarz, P., Burget, L. and Cernocký, J., 2018, June. BUT/Phonexia Bottleneck Feature Extractor. In Odyssey (pp. 283-287).

\bibitem{b27} Abad, A., Ribeiro, E., Kepler, F.N., Astudillo, R.F. and Trancoso, I., 2016, September. Exploiting Phone Log-Likelihood Ratio Features for the Detection of the Native Language of Non-Native English Speakers. In INTERSPEECH (pp. 2413-2417).

\bibitem{bsb} Ravanelli, M., Parcollet, T., Plantinga, P., Rouhe, A., Cornell, S., Lugosch, L., Subakan, C., Dawalatabad, N., Heba, A., Zhong, J. and Chou, J.C., 2021. SpeechBrain: A general-purpose speech toolkit. arXiv preprint arXiv:2106.04624.

\bibitem{b28} Wang, C., Liao, M., Huang, Z., Lu, J., Wu, J., Liu, Y., Zong, C. and Zhang, J., 2023. Blsp: Bootstrapping language-speech pre-training via behavior alignment of continuation writing. arXiv preprint arXiv:2309.00916.

\bibitem{b29} Tang, C., Yu, W., Sun, G., Chen, X., Tan, T., Li, W., Lu, L., Ma, Z. and Zhang, C., 2023. Salmonn: Towards generic hearing abilities for large language models. arXiv preprint arXiv:2310.13289.

\bibitem{b30} Chu, Y., Xu, J., Zhou, X., Yang, Q., Zhang, S., Yan, Z., Zhou, C. and Zhou, J., 2023. Qwen-audio: Advancing universal audio understanding via unified large-scale audio-language models. arXiv preprint arXiv:2311.07919.

\bibitem{b31} Hu, S., Zhou, L., Liu, S., Chen, S., Hao, H., Pan, J., Liu, X., Li, J., Sivasankaran, S., Liu, L. and Wei, F., 2024. Wavllm: Towards robust and adaptive speech large language model. arXiv preprint arXiv:2404.00656.


\end{thebibliography}
\end{document}